\documentclass[a4paper]{article}

\usepackage{INTERSPEECH_v2}

\title{Language Recognition using Time Delay Deep Neural Network}
\name{Mousmita Sarma$^1$, Kandarpa Kumar Sarma$^1$, Nagendra Kumar Goel$^2$}
\address{
  $^1$ Dept. of Elect. and Communication Engineering, Gauhati University, Guwahati, Assam, India\\
  $^2$ Go-Vivace Inc.	McLean VA, USA}
\email{go4mou@gmail.com, kandarpaks@gmail.com, nagendra.goel@govivace.com}

\begin{document}

\maketitle
\begin{abstract}
This work explores the use of a monolingual Deep Neural
Network (DNN) model as an universal background model
(UBM) to address the problem of Language Recognition (LR)
in I-vector framework. A Time Delay Deep Neural Network
(TDDNN) architecture is used in this work, which is trained as
an acoustic model in an English Automatic Speech Recognition (ASR) task. 
%Several different languages are used to train the I-vector system. 
A logistic regression model is trained to classify the I-vectors.
%of 14 different languages of LRE 2007 evaluation. The average cost performance, Cavg , for 3 sec, 10 sec and 30 sec conditions in a GMM-UBM based I-vector system are 20.28\%, 6.23\% and 1.56\% respectively and these reduces to 7.73\%, 1.61\% and 0.24\% respectively in the DNN-UBM based I-vector system. 
The proposed system is tested with fourteen languages with various confusion pairs and it 
can be easily extended to include a new language by just retraining the
last simple logistic regression model. The architectural flexibility is the major
advantage of the proposed system compared to the single DNN
classifier based approach.
\end{abstract}


\section{Introduction}\label{sec:intro}
Language Recognition (LR) refers to a machine based approach through which the identity of the language spoken in a speech sample is authenticated. In recent times, tremendous progress have been made in this regard, which is benefited from technological breakthroughs in related areas, such as signal processing, pattern recognition, cognitive science and machine learning. LR systems are usually categorized by perceptual cues they use, such as the \emph{acoustic-phonetic approach}, the \emph{phonotactic approach}, the \emph{prosodic approach} and the \emph{lexical approach}. However, recent studies have confirmed that acoustic-phonetic and phonotactic features are the most effective language cues and these are the features used typically in the mainstream LR technology \cite{lid1} \cite{LID_review}. 
\par I-vector based approach have been extensively used for LR during the past few years. Such an I-vector system designed for LR task relies on two factors - a front end feature for acoustic representation of the speech signals and a back end model to capture the statistics of distribution of the phonetic events at acoustic feature level. Recent LR experiments have analysed and proved that the shifted  delta  cepstral  (SDC) feature vectors  created  by  stacking  delta  cepstra  computed across multiple speech frames provides best performance in LR \cite{SDC}. This is because they capture the dynamics over a wider range of speech frames than the first and second order MFCC derivatives. On the other hand typically Gaussian Mixture Model (GMM) based Universal Background Model (UBM) is used to collect sufficient statistics (SS) for such I-vector systems. I-vectors are used for scoring directly or used as input to some secondary classifiers. 
\par However, recently Lei at el., 2014 \cite{Lei} have proposed an efficient technique of using Deep Neural Network (DNN)s to compute SS for I-vector systems designed for Speaker Recognition (SR) task. Later a few other works are reported the using of such DNN as a means of computing posterior probability of feature vectors in SR task and observed significant improvement in performance \cite{SR1} \cite{sre10} \cite{UnifiedSRLR}. Such improvements inspires this study to use an English DNN as an UBM to address the LR problem. I-vectors employ subspace techniques to model variations from the UBM. Hence our motivation is that if the DNN-UBM models a normal language, the I-vector shall focus more on the variations in the language. Usually such a DNN is trained for modeling acoustic events like tied triphone state or senones in a supervised way using labeled data. The classes corresponding to senone and posteriors are very accurately computed to predict the senone in such a DNN. Hence  using the posteriors from DNN class, the I-vector system shall be able to differentiate the minute variations of phonemes occurred due to the change of languages. 
\par The DNN architecture used for this work follows the structure of a Time delay DNN (TDDNN) developed by Peddinti et al., 2015 \cite{TDNN}. The architecture is capable to effficiently model the long term temporal dependencies between acoustic events in Automatic Speech Recognition (ASR) task. It uses a very efficient subsampling technique of selecting time steps to compute hidden activations which reduces computation during training. The TDDNN stucture provides advantage compared to the recurrent structures in terms of less computational time, while effectively learning the temporal dynamics of speech signal just like the recurrent structure. 
\par The LR work reported in this paper trains the TDDNN models using around 1800 hours of the English part of Fisher data and uses the TDDNN to compute posterior probability of feature vectors for I-vector extractor training. The I-vector system is trained using data from around 50 different languages. Finally a logistic regression model is trained and tested on top of the I-vectors using the 14 languages of LRE 2007 evaluation. The results reported in this study are compared with a I-vector system which uses a GMM-UBM. However, the major adavantage of the proposed system is its architectural flexibility to include a new language. Keeping rest of the models as it is, just by retraining the simple logistic regression model a new language can be added to the system. However, in single DNN classifier based system adding a new language requires a retraining of the entire DNN model. The work also reported some intermediate results observed during the LR experiments like use of vocal tract length normalization (VTLN) warping and speaker recogntion style features for LR. The results are included for the case when VTLN warping is performed on the acoustic features and how the results varies when that step is removed. Further features used for speaker recognition, like 60 dimensional Mel Frequency Cepstral Coefficients (MFCC) augmented with delta and acceleration, is also used for the work and compared with the standard SDC. 
%Experimental results proves that 60-MFCC provides better result than SDC in 3 sec condition. The same TDNNN architecture has been previously used by Snyder et al., 2015 \cite{sre10} for SR tasks using 60-MFCC acoustic features. Hence the present results also reflects the usability of the TDDNN architecture for an unified SR and LR tasks.
\par The rest of the paper is organized as follows: Section~\ref{sec:baseline} provides a description of the baseline system. Description of DNN-UBM based approach is given in Section~\ref{sec:DNN}. Experimental set up and related discussions are reported in Section~\ref{sec:exp}. Section~\ref{sec:conclusion} concludes the description.
\section{DNN-UBM based approach}
\label{sec:DNN}
A DNN trained as the acoustic model using English speech as part of an ASR, is used as background model for LR in this work. The posteriors of phonetic content like tied triphone states obtained from the output layers of such DNN is used for I-vector system. 
\par In the I-vector model given by Dehak et al. \cite{Ivec}, the total factor $w$ is
defined by the posterior distribution of the $T$ feature vectors of utterence $Y={y_{1},y_{2},\cdots,y_{T}}$ conditioned to the following Baum-Welch statistics,
\begin{equation}\label{eq34}
N_{c}=\sum_{t=1}^{T}P(c|y_{t},\lambda)
\end{equation}
\begin{equation}\label{eq35}
F_{c}=\sum_{t=1}^{T}P(c|y_{t},\lambda)y_{t}
\end{equation}
\begin{equation}\label{eq36}
\hat{F_{c}}=\sum_{t=1}^{T}P(c|y_{t},\lambda)(y_{t}-\mu_{c})
\end{equation}
where, $c$ is the Gaussian index representing $j=1,\cdots,M$ mixture components, $P(c|y_{t},\lambda)$ corresponds to the posterior probability of mixture component $c$ generating the vector $y_{t}$, given UBM $\lambda$ and $\mu_{c}$ is the mean of UBM mixture components $c$. 
\par In practice, the posterior probability of the event that the feature vector $y_{t}$ is generated by the mixture component $c$, is the alignment of feature vector $y_{t}$ represented by the $c^{th}$ Gaussian. In the DNN based I-vector system, the Gaussian index $c$ is replaced by the class index $k$ of DNN and posteriors from the $k^{th}$ DNN class as the $P(c|y_{t},\lambda)$s in eq.s~\ref{eq34} to ~\ref{eq36}. Basically the posteriors of the $c^{th}$ Gaussian which is used as the alignments of a feature vector is represented by the senones predicted by the DNN \cite{Lei}. 
\par The DNN architecture used in this study is a multisplice TDDNN \cite{TDNN} which is capable of capturing long term temporal dependency of phonetic contents. To represent long term spectro-temporal dynamics of the speech signal various approaches are proposed in earlier literature. Feature representation like TRAP \cite{TRAP}, HATS \cite{HATS} etc. can be used with standard feed forward model ANN. On the other hand Recurrent Neural Network (RNN) along with Long Short Term Memory (LSTM) architectures are used for modeling of temporal context \cite{rnn} in deep architectures. But the TDDNN architecture provides some relative advantages over such approaches in terms of computational complexity, training time and parallelization, by utilising a subsampling technique. Here, a narrow temporal context is provided to the first layer and increasingly wide contexts are available to the subsequent hidden layers. The result is that higher levels of the network are able to learn greater temporal relationships. Due to selective computation of time steps such sub-sampling scheme reduces the overall necessary computation during the forward and backward pass computation \cite{TDNN}. The final TDDNN has six layers. If $t$ is some frame, at layers $0$, $1$, $3$ and $4$, frames $[t-2, t+2]$, $[t-2, t+1]$, $[t-3, t+3]$ and $[t-7, t+2]$ respectively are spliced together to form the subsampling configuration. The DNN configuration can be summarised as follows: P-norm activation function is used in the hidden layers and output is a softmax layer. The network is trained using Natural Gradient for Stochastic Gradient Descent (NG-SGD) algorithm \cite{NG}. Initial effective learning rate is 0.0015 and final effective learning rate is 0.00015. The network is trained for 6 number of epochs. 
\section{Full covariance GMM-UBM based approach}
\label{sec:baseline}
A Full Covariance GMM-UBM \cite{sre08} is used to train the I-vector extractor for LR. In such a system, the GMM-UBM plays the role of providing the SS required to train the I-vector model. Since collection of statistics with full covariance GMM is computationally expensive, hence a simplified approach where a tandem of coherent diagonal-covariance and full-covariance models was proposed in \cite{sre08}. Basically in UBM training first a diagonal covariance UBM is trained. Then, one iteration of the expectation maximization (EM) is run with fixed means and mixture weights, to obtain a full-covariance model. For any utterance $X$, first a small number of Gaussians are selected on the basis of diagonal model, and then posteriors are computed using the full version of Gaussians with the same means. The performance of this GMM-UBM based system is compared with the proposed DNN-UBM based approach. The system uses same dataset for training I-vector extractor and logistic regression model which are used for DNN-UBM based approach.
\section{Experimental Set up, Results and Analysis}
\label{sec:exp}
\begin{figure*}[t]
\centering
\includegraphics[width=15.0cm, height=5.0cm]{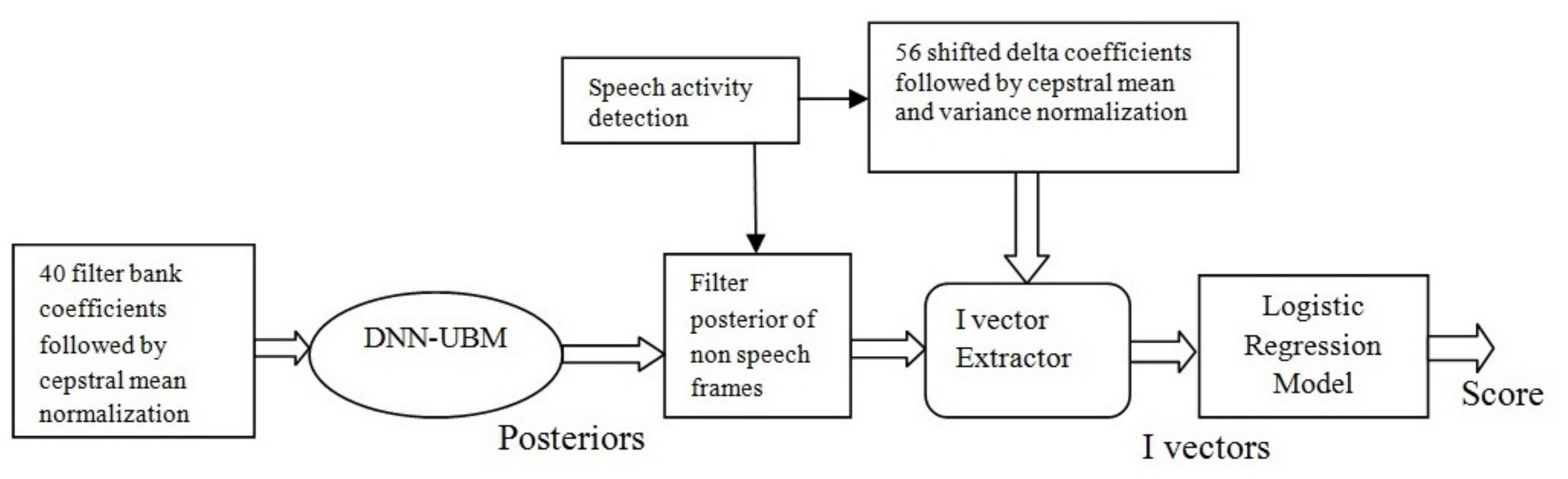}
\caption{DNN based Langauge Recognition (LR) Framework}\label{FigLID}
\end{figure*}

All experiments are performed in the Kaldi Speech Recognition Toolkit \cite{kaldi}. The scripts for the DNN-UBM based LR framework and various experiments reported here are now available as the LRE07/v2 example of the Kaldi code repository (https://github.com/kaldi-asr/kaldi/blob/master/egs/lre07/v2/run.sh). Figure~\ref{FigLID} shows the process logic of the system. The experimiental results are reported in terms of classification Error Rate (ER) and Average Cost ($C_{avg}$) \cite{LRE2007}.
\subsection{Database}
\label{sec:data}
%For the entire system three sets of data are required. First the development data to build the DNN-UBM, secondly the training data to train the UBM (for baseline system) and the I-vector extractor and lastly the testing data to evaluate the performance. Here 
The DNN is trained using the 1800 hours of  Fisher English. The SRE 2008 training set, CALLFRIEND Vietnamese, Tamil, Japanese, Hindi, German, Farsi, French, Standard Arabic, Korean, Mainland Chinese Mandarin, Taiwan Chinese Mandarin, Caribbean Spanish, Non-Caribbean Spanish, LRE 1996, LRE 2003, LRE 2005, LRE 2009 and LRE 2007 supplimentary data are used here for training the I-vector extractor. Thus the I-vector extractor training data has around 50 languages and accents. The LRE 2007 evaluation data is used for evaluating the performance of the system.  
\subsection{Acoustic feature extraction and speech activity detection}
\label{sec:feature}
The acoustic features are collected from 20 ms window with a overlap of 10 ms. Low energy  dither  is  added  to  the  windowed signal  to  avoid  runs  of digital zeros.  Per-file vocal tract length normalization (VTLN) warps are applied. SDC  features  are  extracted  using   the   conventional   7-1-3-7   scheme and the 7 static cepstra   are   appended   producing   a   56-dimensional feature vector. This is followed by non-speech frame filtering using energy based speech activity detection and cepstral mean normalization of 3 sec window. For experimental purpose, 19 MFCC augmented with energy, 20 delta and 20 acceleration coefficients, creating 60-dimensional MFCCs are also computed and evaluated using the system separately.
\subsection{GMM-UBM training}
\label{ssec:GMMUBM}
Initially, one diagonal UBM is trained with a small subset of training data (5000 utterances) with 2048 mixture components using the variance equal to the global variance of the features and the mean equal to the mean of distinct and randomly chosen frames. This UBM is used to train a full covariance UBM with same 2048 mixture components with another subset of development data (10000 utterances)  which is again used to initialize the final full covariance UBM using the entire training data. I-vector extractor is trained utilizing both full covariance and diagonal covariance in a computationally efficient manner as described in Section~\ref{sec:baseline}.
\subsection{Replacement of GMM-UBM with DNN-UBM}
\label{ssec:DNN-UBM}
In order to replace the GMM-UBM with a DNN-UBM another set of acoustic features are required. To compute posteriors from the DNN, it uses a special set of high-resolution MFCC features. This is formed by all 40  filterbank coefficients without cepstral truncation extracted from frame-length of 25ms with 10 ms shift. The intention is to provide raw speech features of spectrogram to the DNN, since the deep architectures have inherent capability of learning features from the input signal. Cepstral mean subtraction is performed over a window of 6 seconds. The non speech frames are not filtered beforehand. The posteriors are computed from all frames to maintain actual temporal context and posteriors related to non speech frames are later filtered out. However, to train the T-matrix of I-vector system it uses the standard 56-dimensional SDC features.
\subsection{I-vector extractor training}
Two separate I-vector extractors are trained. First using the full covariance GMM-UBM, which provides the GMM-UBM I-vector based system for comparison. Secondly, the DNN-UBM based I-vector extractor is trained using posteriors computed from DNN. This I-vector extractor is initialized using a separately trained supervised GMM \cite{sre10}. This GMM is supervised because DNN posteriors are used for training. However supervised GMM is used only to initiate the I-vector extractor, but the training of the T-matrix uses the DNN posteriors. The 40 filterbank features are used to compute the posteriors from the DNN. But SDC features are used for T-matrix training. The posteriors related to non speech frames are later filtered out using the speech activity detection information used for the LR features. All I-vectors are 600 dimensional. Two sets of I-vectors for all the testing and training data are computed using both this GMM based I-vector extractor and DNN-based I-vector extractor.
\subsection{Logistic regression based classification}\label{sec4_6}
The I-vectors of the 14 languages namely Arabic, Bengali, Chinese, English, Farsi, German, Hindustani, Japanese, Korean, Russian, Spanish, Tamil, Thai and Vietnamese are used for training the logistic regression model. This logistic regression model is finally used for LR evaluation. Posterior probabilities for the I-vectors of the LRE 2007 evaluation datasets are computed and these posteriors are used for scoring. 

\subsection{Evaluation of the System and Discussion} 
\begin{table}[t]
\centering
\caption{Error rate (ER) and $C_{avg}$}\label{lre_result}
\begin{tabular}{lllll  }
\hline
\multicolumn{5}{c}{Part (a): Scoring of GMM-UBM based }\\
\multicolumn{5}{c}{system with SDC features and VTLN warping} \\
\hline
Duration (sec)            &   3 sec     & 10 sec       &  30 sec  &  Average  \\
\hline
   ER  (\%)               &   32.58      &9.87         &2.92      &15.12         \\
%\hline
   $C_{avg}$  (\%)        &    20.28     &6.23         &1.56      &9.36       \\
\hline
\multicolumn{5}{c}{Part (b): Scoring of DNN-UBM based system } \\
\multicolumn{5}{c}{with SDC features and VTLN warping} \\
\hline
Duration (sec)         &   3 sec     & 10 sec       &  30 sec  &  Average  \\
\hline                          
 ER (\%)               &   12.98     & 2.69         &  0.42     &  5.36          \\
$C_{avg}$ (\%)         &   7.73      & 1.61         &  \textbf{0.24}     &  3.19  \\
\hline       
\multicolumn{5}{c}{Part (c): Scoring of DNN-UBM based system } \\
\multicolumn{5}{c}{with SDC features (VTLN skipped)} \\
%\hline
Duration (sec)         &   3 sec     & 10 sec       &  30 sec  &  Average  \\
\hline
 ER (\%)               &   20.81     &4.68          &1.02      &8.84         \\
$C_{avg}$ (\%)         &  12.33     &2.66          &0.48       &5.16 \\
\hline
\multicolumn{5}{c}{Part (d): Scoring of DNNN-UBM system} \\
\multicolumn{5}{c}{speaker recognition style 60-MFCC features (VTLN skipped)} \\
\hline
Duration (sec)         &   3 sec     & 10 sec       &  30 sec  &  Average  \\
\hline
   ER (\%)             &   18.81     & 5.10         &  1.16      &8.36      \\
   $C_{avg}$ (\%)      &   11.87    &3.31        &  0.83         & 5.34 \\
\hline
\end{tabular}
\end{table}
The system is evaluated using the 14 language targets as outlined in Section~\ref{sec4_6}, which are specified in NIST LRE 2007 evaluation. All target languages also serve as the non-target (alternative hypothesis) languages, which forms the target/non-target language pairs for evaluation. The speech segments are extracted from one side of telephone conversation at standard 8-bit, 8-kHz, $\mu$-law format. Basic pair-wise LR performance are computed for all target/non-target language pairs at 3 sec, 10 sec and 30 sec testing conditions and finally average cost performance $C_{avg}$ \cite{LRE2007} is computed. 
%All experiment were performed on an Intel x86-64 machine with 8 CPU, 4 GHz CPU speed and 32 GB RAM. However DNN training and posteriors computations are performed in CUDA using a NVIDIA-SMI 358.16 GPU. 
\par The DNN-UBM based I-vector extractor is compared with GMM-UBM based I-vector extractor. Part (a) and part (b) of Table~\ref{lre_result} represents the ER and $C_{avg}$ for GMM-UBM and DNN-UBM systems. The GMM-UBM based system achieves $C_{avg}$ of \%, \% and \% for 3 sec, 10 sec and 15 sec conditions and these numbers reduces to 7.73\%, 1.61\% and 0.24\% for the proposed DNN-UBM based system. The results clearly state that use of DNN-UBM to collect SS for I-vector extractor provides a significant improvement in individual $C_{avg}$ compared to GMM-UBM for all three test conditions. 
Further it has been observed that the performance of the proposed TDDNN-UBM system is relatively better than the results reported in \cite{UnifiedSRLR} using SDC features in DNN and I-vector framework ($C_{avg}$ of  19.5\%, 8.21\% and 4.00\% for 3 sec, 10 sec and 15 sec conditions respectively), using bottleneck features DNN  and I-vector framework ($C_{avg}$ of 18.2\%,  7.71\% and 3.79\% for 3 sec, 10 sec and 15 sec conditions) in LRE 2011 evaluation. 
%Fig~\ref{Cavg_GMM_TDDNN} provides a comparison of $C_{avg}$ for 3 sec, 10 sec and 30 sec conditions.  
The reason for this improvement is perhaps the use of an uniform UBM which contributes all variations to the I-vector, hence the I-vector is more pure with respect to language signal. The current study could not provide a comparison using LRE 2011 database since the database is not yet available publicly.  Further, another advantage of the proposed approach is that, its easy to extend the system to a new language even if limited audio is available in a new unknown language, by just doing the final logistic regression step. The work described in \cite{dnn1} and \cite{dnn2} presents results using DNN based feedforward classifier and LSTM-RNN based classifier. However those systems are evaluated choosing 8 unidentified languages out of 23 languages in LRE09. But the present results use the 14 languages of LRE 2007 evaluation and applied no filtering based on the amount of data available or language pair confusability. The current architecture allows easy addition of a new language because only the last layer of ligistic regression can be retrained, and the ivector-extractor retraining can be skipped in most cases.
\par Part (c) of Table~\ref{lre_result} represents the performance difference if we skip the VTLN warping of speech signal. This study also experimented with 60-MFCC features for LR. The I-vector extractor is trained using 60 dimensional MFCC-delta-acceleration features (which are standardized in speaker recognition task) and DNN posteriors. Evaluation results are shown in part (d) of Table~\ref{lre_result}. Individual comparison reflects that for 30 sec condition MFCC provides similar performance, however for short duration like 3 sec condition SDC performs better.

\section{Conclusion}
\label{sec:conclusion}
The work reported here has integrated a TDDNN based UBM in an I-vector based LR system replacing the GMM-UBM to compute posterior distribution of feature vectors. The DNN used here is trained as an acoustic model in a monolingual English ASR. 
%Our motivation was that, since I-vectors employ subspace techniques to model variations from the UBM, hence if the DNN-UBM models a normal language, the I-vector shall focus more on the variations in the language. 
The expermental results have shown significantly low average cost performance compared to the GMM-UBM based system. The current set up uses 14 languages for evaluation. However in order to include a new language to the system, the final Logistsic regression model is the only component which is required to retrain. This architectural flexibility is another major advantage of the proposed sysetm.
%\begin{figure}
%\centering
%\includegraphics[width=9cm]{Cavg_GMM_TDDNN.pdf}
%\caption{Comparative analysis of $C_{avg}$ in GMM-UBM and TDDNN-UBM system}\label{Cavg_GMM_TDDNN}
%\end{figure}
%The system takes advantage of capability of the TDDNN architecture to capture long term temporal context of feature space and to compute posteriors accurately due to supervised learning unlike a GMM-UBM. 
%\vfill\pagebreak
%\begin{figure}
%\centering
%\includegraphics[width=9cm]{Cavg_SDC_MFCC.pdf}
%\caption{Comparative analysis of $C_{avg}$  of SDC and 60-MFCC features in TDDNN-UBM system}\label{Cavg_SDC_MFCC}
%\end{figure}
\section{Acknowledgement}
We would like to thank David Snyder and Daniel Povey of Center for Language and
Speech Processing, The Johns Hopkins University, Baltimore, MD, USA. for their helpful comments and review. 

\bibliographystyle{IEEEtran}

\bibliography{mybib}

% \begin{thebibliography}{9}
% \bibitem[1]{Davis80-COP}
%   S.\ B.\ Davis and P.\ Mermelstein,
%   ``Comparison of parametric representation for monosyllabic word recognition in continuously spoken sentences,''
%   \textit{IEEE Transactions on Acoustics, Speech and Signal Processing}, vol.~28, no.~4, pp.~357--366, 1980.
% \bibitem[2]{Rabiner89-ATO}
%   L.\ R.\ Rabiner,
%   ``A tutorial on hidden Markov models and selected applications in speech recognition,''
%   \textit{Proceedings of the IEEE}, vol.~77, no.~2, pp.~257-286, 1989.
% \bibitem[3]{Hastie09-TEO}
%   T.\ Hastie, R.\ Tibshirani, and J.\ Friedman,
%   \textit{The Elements of Statistical Learning -- Data Mining, Inference, and Prediction}.
%   New York: Springer, 2009.
% \bibitem[4]{YourName17-XXX}
%   F.\ Lastname1, F.\ Lastname2, and F.\ Lastname3,
%   ``Title of your INTERSPEECH 2017 publication,''
%   in \textit{Interspeech 2017 -- 18\textsuperscript{th} Annual Conference of the International Speech Communication Association, August 20?24, Stockholm, Sweden, Proceedings, Proceedings}, 2017, pp.~100--104.
% \end{thebibliography}

\end{document}